\documentclass[
 reprint,
 amsmath,amssymb,
 aps,
pra,
reprint, superscriptaddress, amsmath, amssymb, aps, pra,]{revtex4-2}

\usepackage{tikz}
\usepackage{graphicx}% Include figure files
\usepackage{dcolumn}% Align table columns on decimal point
\usepackage{bm}% bold math
\usepackage{hyperref}% add hypertext capabilities
\usepackage{mathtools}
\usepackage[normalem]{ulem} % evita alterar o estilo de \emph

\DeclarePairedDelimiterX\braket[2]{\langle}{\rangle}{#1\,\delimsize\vert\,\,#2}

\usepackage{xcolor}

\begin{document}

\preprint{APS/123-QED}

\title{Tailoring resonator geometry for astigmatism compensation with structured light in birefringent optical cavities} %Optimal geometrie

%\author{J. C. de Carvalho, Jr.$^{1}$}\email{Corresponding author: juscelino_campos@hotmail.com}
 % Autor correspondente
%\author{R. F. Barros$^{2}$}
%\author{A. Z. Khoury$^{1}$}
%\author{C. E. R. Souza$^{1}$}
%\author{G. B. Alves$^{1}$}
%\affiliation{$^{1}$Instituto de Física, Universidade Federal Fluminense, CEP 24210-340, Niterói-RJ, Brazil}
%\affiliation{$^{2}$Instituto de Física, Universidade de São Paulo, CEP 05508-090, São Paulo-SP, Brazil}

\author{J. C. de Carvalho Jr.}
\email[Corresponding author: ]{juscelino.ccjr@gmail.com}
\affiliation{Instituto de Física, Universidade Federal Fluminense, CEP 24210-340, Niterói-RJ, Brazil}

\author{R. F. Barros}
\affiliation{Instituto de Física, Universidade de São Paulo, CEP 05508-090, São Paulo-SP, Brazil}

\author{A. Z. Khoury}
\affiliation{Instituto de Física, Universidade Federal Fluminense, CEP 24210-340, Niterói-RJ, Brazil}

\author{C. E. R. Souza}
\affiliation{Instituto de Física, Universidade Federal Fluminense, CEP 24210-340, Niterói-RJ, Brazil}

\author{G. B. Alves}
\email[Corresponding author: ]{gbie@id.uff.br}
\affiliation{Instituto de Física, Universidade Federal Fluminense, CEP 24210-340, Niterói-RJ, Brazil}

%\date{\today}

\begin{abstract}
In this work, we experimentally investigate the astigmatism induced by a birefringent crystal inside an optical cavity and demonstrate a geometric compensation strategy for structured modes carrying orbital angular momentum (OAM). We show that the astigmatism observed in the semimonolithic configuration can be fully compensated by introducing a third mirror to fold the cavity, thereby exploiting the geometric astigmatism arising from oblique incidence on the curved mirror. By performing a study of the resonator geometry, both theoretical and experimental results confirm that this approach enables the preservation of OAM between the pump beam and the converted signal and idler modes, providing an effective design strategy for birefringent optical cavities.
\end{abstract}

\maketitle

\section{\label{sec:level1}INTRODUCTION}

The production of structured light modes has become a prominent topic in the scientific literature due to their ability to overcome the limitations of conventional optical beams \cite{forbes_structured_2021}. By offering new degrees of freedom—such as Orbital Angular Momentum (OAM) \cite{Padgett:17}—structured light changed the paradigm of data transmission and material manipulation. This versatility enables high-impact applications, including high-capacity optical communications via spatial mode division multiplexing \cite{doi:10.1126/science.1237861}, precise optical tweezers for biological manipulation \cite{PhysRevLett.131.163601}, customized laser materials processing \cite{PhysRevLett.110.143603}, and super-resolution imaging \cite{Torok:04}.

There are two main approaches to generate these modes. Extracavity beam shaping transforms a standard Gaussian beam outside the laser cavity using components such as Spatial Light Modulators (SLMs) or spiral phase plates. Although highly flexible, this method suffers from high energy losses and low power-damage thresholds. Conversely, intracavity generation forces the laser oscillator to produce the structured mode directly at the source. This is achieved through gain shaping via off-center pumping \cite{sheng_tunable_2026} or controlled astigmatism \cite{Lu:13}, such as tilting mirrors or inserting cylindrical lenses \cite{PhysRevApplied.14.044048}. This geometric symmetry-breaking suppresses competing modes, resulting in high modal purity, superior energy efficiency, and high-power resistance.

When it comes to nonlinear interactions in an optical cavity, there is an additional challenge coming from the anisotropy of the nonlinear crystal~\cite{martinelli_orbital_2004}. Even when the interacting fields have the same polarization, the anisotropy induces an astigmatism effect, breaking the cylindrical symmetry of the optical resonator. This issue prevents general structured light modes, including, but not limited to OAM modes, from being eigenmodes of birefringent resonators. Examples of intracavity nonlinear interactions are optical parametric oscillation, which is paramount for ultrafast optics~\cite{zadeh-2014} as well as for the generation of squeezed light fields~\cite{SCHNABEL20171}, as well as quantum frequency conversion~\cite{Kumar:90} and cavity-enhanced second-harmonic generation~\cite{Ou:92}. In general, second-order nonlinear media only achieve phase matching due to birefringence, even in periodically poled crystals with non-critical phase matching.

Optical parametric oscillators (OPO) are the primary sources used to produce squeezed light to reduce quantum noise below the standard vacuum limit — essential for improving the sensitivity of gravitational wave detectors \cite{SCHNABEL20171} — as well as to generate spatially structured entangled states \cite{PhysRevLett.113.170501,Cai:18}. When integrated with structured light engineering, OPOs provide a robust framework to down-convert complex spatial profiles and orbital angular momentum directly at the source, paving the way for high-capacity quantum communications, advanced optical manipulation, and high-power laser material processing. Moreover, the generation of squeezed light in high-order structured light modes has been a topic of intense research, especially for metrology purposes. In references \cite{PhysRevLett.113.170501,Cai:18}, astigmatism compensation for structured light modes in an OPO is achieved by inserting a half-wave plate and an orthogonally rotated nonlinear crystal into the cavity, thereby requiring extra optics inside the resonator.

Here we propose the use of an astigmatic cavity to compensate for the intrinsic astigmatism generated by the anisotropic nonlinear crystal. It shows how the geometric engineering of the optical cavity allows the first-order Laguerre modes of ordinary and extraordinary polarizations to resonate simultaneously. This methodology introduces a promising approach for generating non-classical states in an (type-II) optical parametric oscillator, overcoming the astigmatism asymmetry that typically penalizes one of the polarizations and would prevent clean resonance of higher-order modes.

This paper is organized as follows. In Sec.\ref{Sec:theo_model}, we present the theorical model, in which we discuss the round-trip Gouy phase in a high-finesse linear ressonator and the effect of the astigmatism introduced by the crystal anisotropy. We begin by analyzing the astigmatic splitting of Hermite-Gaussian modes in a semimonolithic cavity, showing that this effect can be mitigated by increasing the mirror radius of curvature $R$, albeit at the cost of increasing the Rayleigh length and compromising beam focusing. To preserve this parameter without disrupting mode-matching, we propose a folded three-mirror cavity setup.  In Sec.~\ref{Sec:exp_res}, we present the experimental results for high-finesse resonators: we show the effect of astigmatism in a semimonolithic cavity; we demostrate that the folded cavity compensates for the crystal astigmatism, anabling OAM conservation; and, finally, we show that tuning the radius of curvature of a linear also enables simultaneous resonance in both polarizations of the crystal. In Sec.~\ref{Sec:conclusion}, we present the conclusions. The Appendices contain the theorical derivations that support the results presented.

\section{THEORETICAL MODEL}\label{Sec:theo_model}

\subsection{Astigmatic structured light}

The Gouy phase plays an important role in describing the transverse modes that resonate in the cavity \cite{yariv2007photonics}. For an astigmatic Hermite-Gaussian ($HG_{m,n}$) mode, the Gouy phase is given by \cite{BEIJERSBERGEN1993123}:

\begin{equation}
\Phi = \left(m + \frac{1}{2}\right)\phi_{x} + \left(n + \frac{1}{2}\right)\phi_{y},
\end{equation}
where $m$ and $n$ are the transverse mode indices, $\phi_{x}$ and $\phi_{y}$ are the Gouy phases accumulated along the transverse plane ($x, z$), and the transverse plane ($y, z$). Here we consider the beam propagating along the $z$ axis. Expressions for $\phi_{x}$ and $\phi_{y}$ are given by:

\begin{align}
\phi_{x} = \arctan\left(\frac{z-z^{x}_0}{z_{R_x}}\right), \quad
\phi_{y} = \arctan\left(\frac{z-z^{y}_0}{z_{R_y}}\right),
\end{align}
where, $z^{x}_0$ and $z^{y}_0$ are the waist positions, while $z_{Rx}$ and $z_{Ry}$ are the Rayleigh lengths in the transverse directions. The astigmatism effect arises when $z_{Rx}\neq z_{Ry}$, leading to a Gouy phase that depends on $m$ and $n$ independently rather than on the total mode order $m+n$~\cite{PhysRevA.103.023511}. As a consequence, Laguerre-Gaussian ($LG$) modes, which are superpositions of $HG$ modes of the same order and require cylindrical symmetry, are not self-similar in astigmatic systems and therefore cannot be eigenmodes of astigmatic cavities. 

The astigmatism effect arises from cylindrical lenses, oblique reflections from spherical mirrors, or propagation in birefringent media~\cite{martinelli_orbital_2004}. Although birefringence refers to polarization-dependent refraction, the polarization of a focused laser beam varies across the angular spectrum, leading to astigmatism even when the input field is initially polarized along one of the crystal axes. This phenomenon is discussed in Ref.~\cite{martinelli_orbital_2004} and summarized in Appendix~\ref{Sec:app_A}. While birefringent-induced astigmatism is generally negligible in free-space propagation, in cavity-enhanced nonlinear interactions, such as optical parametric oscillation, quantum frequency conversion, and second-harmonic generation, its cummulative effect over multiple round trips is sufficient to prevent the resonance of $LG$ modes, as we discuss in the following sections.

\subsection{Conditions for OAM preservation in a semimonolithic cavity}\label{Sec:condition_semi}
 
The effect of astigmatism induced by a birrefringent crystal on beams carrying orbital angular momentum was experimentally observed in a linear cavity by Martinelli et al.~\cite{martinelli_orbital_2004}, where $LG_{0}^{\pm1}$ mode was prepared as the pump of an OPO. The crystal anisotropy acts differently on each polarization, selecting the resonant modes in the cavity: for one polarization, the phase splitting between the $HG_{10}$ and $HG_{01}$ modes exceeds the cavity linewidth, preventing OAM conservation, while for the other polarization this separation remains below the linewidth.

\begin{figure}[h]
\centering
\includegraphics[width=0.49\textwidth]{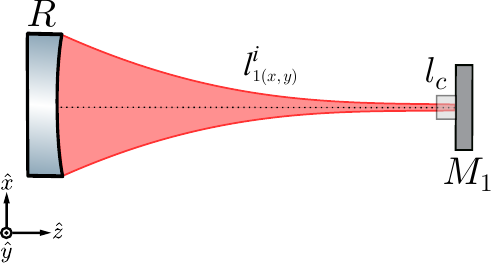}
\caption{Schematic of the semimonolithic cavity. $M_1$ are flat mirror placed at the beam waist positions, $R$ is the curved mirror radius of curvature, and $l_{1(x,y)}^i$ are the cavity arm effective, where $i=\{o,e\}$ is ordinary ($o$) and extraordinary ($e$) polarizations, and $l_{c}$ is the legth of the crystal.}
\label{fig:semi_monolitica}
\end{figure}

In this section, we investigate this effect and the conditions required for OAM conservation in the context of a semimonolithic cavity, as illustrated in Fig.\ref{fig:semi_monolitica}.

The round-trip accumulated Gouy phase \cite{PhysRevA.103.023511} is given below by:

\begin{equation}
    \phi^{i}_{j} = 2\arctan \left(\frac{l^{i}_{1, \,j}}{z^{i}_{R,\,j}}\right),\quad z^{i}_{R,\,j} = \sqrt{l^{i}_{1, \,j}\left(R-l^{i}_{1, \,j}\right)},
\end{equation}
where $j = (x, y)$ denotes the transverse directions, $l^{i}_{1,\, j}$ is the effective optical path length given by Eq.~\eqref{eq:length_eff}, where $i=\{o,e\}$ stands for ordinary ($o$) and extraordinary ($e$) polarizations, and $z^{i}_{R,j}$ is the corresponding Rayleigh length.

We define the Gouy phase difference between the $HG_{10}$ and $HG_{01}$ modes for each polarization as 
\begin{align}
\delta\Phi^{o} &= \Phi^{o}_{10} - \Phi^{o}_{01}, \label{eq:dif_one_order_o} \\[6pt]
\delta\Phi^{e} &= \Phi^{e}_{10} - \Phi^{e}_{01}. \label{eq:dif_one_order_e}
\end{align}

The condition for OAM preservation is satisfied when the phase separation $\delta \Phi^{i}$ is smaller than the cavity linewidth $\delta \phi$ (full width at half maximum) for both polarizations simultaneously, ensuring that the $HG_{10}$ and $HG_{01}$ modes are transmitted within a single resonance peak. 

This necessary condition can be stated as 

\begin{equation}\label{eq:condition_resonance}
|\delta\Phi^{o}|,\, |\delta\Phi^{e}| \leq \delta \phi,
\end{equation}
where $\delta\Phi^{i}$ depend on the resonator geometry and $\delta\phi$ depends on the reflectivity of the cavity mirrors.

Considering the most stringent scenario of a cavity with high-reflectivity mirrors, with intensity reflectivity $\mathcal{R} \to 1$, the finesse parameter can be approximated as $\mathcal{F} \approx \pi \sqrt{\mathcal{R}}/(1 - \mathcal{R})$ ~\cite{yariv1989quantum, suter2014calculation}. The term $\sqrt{\mathcal{R}}$ is the field amplitude attenuation factor per round-trip, arising from the intensity reflectivity $\mathcal{R}$ at each reflection, which we consider equal for all mirrors. 
For an optical resonator, the phase linewidth can be expressed as \cite{yariv1989quantum}

\begin{equation}
\delta\phi = \frac{2\pi}{\mathcal{F}} = \frac{2(1-\mathcal{R})}{\sqrt{\mathcal{R}}}.
\end{equation}

Considering overall losses on the order of $1\%\, (\mathcal{R}=99\%)$, including not only the mirror reflectivities but also the transmission at each facet of the nonlinear medium, we obtain $\delta \phi \simeq 20$ mrad as an estimate for the cavity linewidth. 
Since the birefringent medium induces a phase separation between the ordinary $\delta \Phi^{o}$ and extraordinary $\delta \Phi^{e}$ polarizations of the crystal, we ensure that the separation remains below the cavity linewidth, thereby preserving the simultaneous resonance of modes of the same order in both polarizations.
\begin{figure}[h]
\centering
\includegraphics[width=0.492\textwidth]{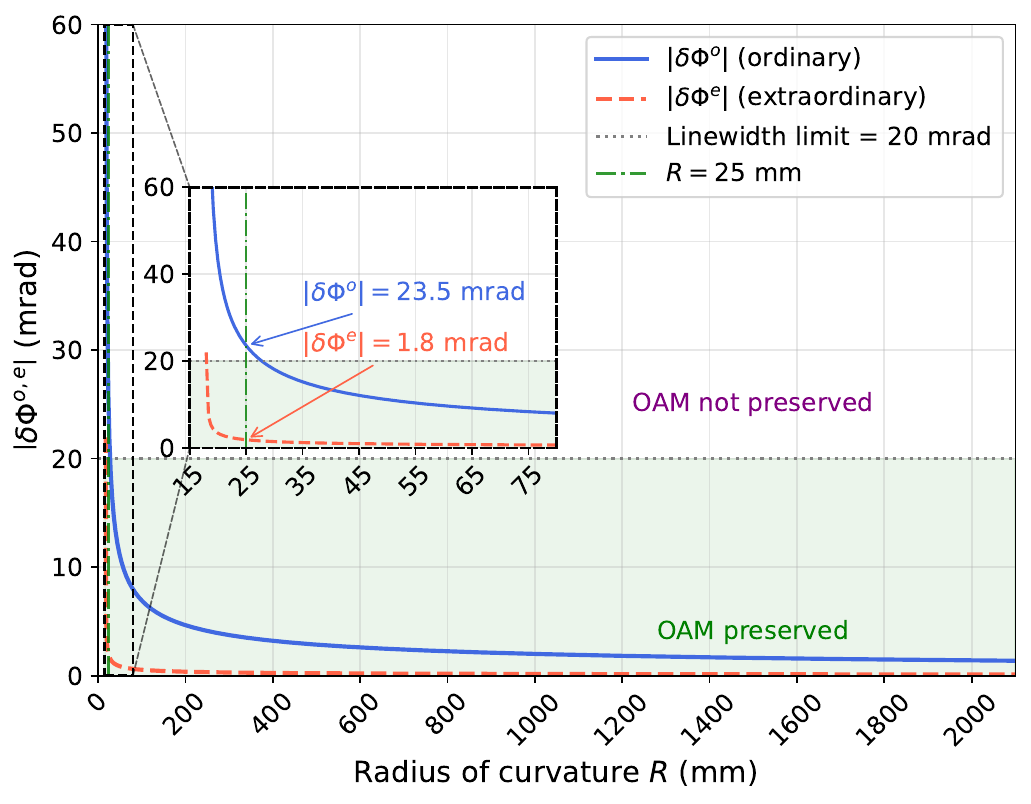}
\caption{The Gouy phase separation $|\delta\Phi^{o,e}|$ between the $HG_{10}$ and $HG_{01}$ modes as a function of the mirror radius of curvature $R$ for a semimonolithic cavity with $l_{1} = 20$ mm and $l_{c} = 5$ mm. The dashed gray indicates the typical linewidth limit of 20 mrad for high-reflectivity. The insert shows a zoom of the region $R \leq  80$ mm, highlighting the experimental point $R = 25$ mm.}%          \textcolor{red}{Gouy phase separation $\delta\Phi^{i}_{j}$ between the $HG_{10}$ and $HG_{01}$ modes as a function of the mirror radius of curvature $R$, for a semimonolithic cavity with $l_{c}$ = 5 mm. For \textbf{(a)} $l_{1}$ = 20 mm, corresponding to the $R$ = 25 mm design point. \textbf{(b)} $l_{1}$ = 46 mm, corresponding to $R$ = 2000 mm. Black solid and red dashed curves show $|\delta\Phi^{o}|$, and $|\delta\Phi^{e}|$, respectively. The dotted gray line marks the typical linewidth limit of 20 mrad for high-reflectivity mirrors. The shaded green region indicates where this condition is satisfied (OAM preserved), while the region above the limit is labeled not preserved. The filled circles and vertical dash-dotted lines mark the respective experimental design points.}}
\label{fig:gouy_R}
\end{figure}

The phase separation between the first-order modes is given by Eqs.~\eqref{eq:dif_one_order_o} and~\eqref{eq:dif_one_order_e}. As $R$ increases, the phase separation continuously decreases as illustrated in Fig.~\ref{fig:gouy_R} until it falls below the cavity linewidth. 
In the limit $R \gg l^{i}_{1,\,j}$, the phase approaches $\Phi^{i}_{j} \approx 2\sqrt{l^{i}_{1, \,j}\,/R}$; thus, the phase separation decreases slowly specifically, reducing $\delta\Phi^{i}$ by a factor of 10 would require increasing $R$ by a factor of 100. Conversely, increasing $R$ also increases the Rayleigh length, causing the mode wavefront structure to approach that of a plane wave characterized by a large waist and weak focusing, which can raise the oscillation threshold in the case of an spontaneus nonlinear proccess.
\subsection{Effect of the Resonator Geometry on preservation OAM}

\begin{figure}[h]
\centering
\includegraphics[width=0.45\textwidth]{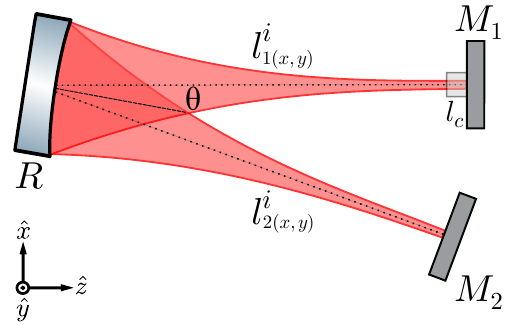}
\caption{Schematic of the folded cavity. $M_1$ and $M_2$ are flat mirrors placed at the beam waist positions, $R$ is the curved mirror radius of curvature, and $l_{1, \,j}^{i}$ and $l_{2, \,j}^{i}$ are the cavity arm  effective optical path length where $i = (o, e)$ denotes the ordinary and extraordinary
polarizations, and these lengths are defined as the distances from each flat mirror to the curved mirror and where $l_{c}$ is the legth of the crystal. The angle $\theta$ is formed between each arm and the optical axis of the curved mirror, such that the total opening angle between the two arms is $2\theta$.}
\label{fig:folded_cavity}
\end{figure}

In order to compensate for the astigmatism introduced by the birefringent crystal in the resonator \cite{martinelli_orbital_2004}, we theoretically modeled an optical cavity capable of minimizing the effects of the crystal’s birefringence on beams with orbital angular momentum. The architecture of this cavity is illustrated in Fig.~\ref{fig:folded_cavity} and consists of an arrangement of three mirrors: two plane mirrors $M_{1}$ and $M_{2}$, positioned at the beam waists, and a curved mirror $R$ outside its optical axis. This geometric arrangement exhibits astigmatism, since the oblique incidence on the curved mirror defines two propagation planes with independent focal lengths: the tangential plane ($xz$) and the sagittal plane ($yz$)~\cite{Penzkofer1996KerrLens}. 

The oblique incidence of the beam on the off-axis curved mirror introduces two distinct focal lengths, given, respectively, by \cite{siegman1986lasers, Massey:69}:
\begin{align}
    f_x &= \frac{R\cos(\theta)}{2},\label{eq:tan} \\[6pt]
    f_y &= \frac{R}{2\cos(\theta)},\label{eq:sag}
\end{align}
where $R$ is the radius of curvature of the mirror and $\theta$, in our case, is the angle formed between the incident beam and the optical axis of the curved mirror.
The result obtained in Appendix~\ref{APsub:3} provides the accumulated round-trip Gouy phase for an $HG_{mn}$ mode which is given by:

\begin{equation}
\begin{aligned}
\Phi^{i}_{m n} &= 2\left(m + \frac{1}{2}\right) \arctan \left( \frac{z_{Rx}^{i}}{f_x - l_{1,x}^{i}} \right) \\[3pt]
&+2 \left(n + \frac{1}{2}\right) \arctan \left( \frac{z_{Ry}^{i}}{f_y - l_{1,y}^{i}} \right)\label{eq:gouy_semi0},
\end{aligned}
\end{equation}
where $i = (o, e)$ denotes the ordinary and extraordinary polarizations, respectively, $l^i_{1x}$ and $l^i_{1y}$ denote effective cavity lengths, as defined in Eq.~\eqref{eq:length_eff}.%, and $f=R/2$ is the focal length of the curved mirror with curvature radius R.

From Eq.~\eqref{eq:gouy_semi0}, the Gouy phase difference between the $HG_{10}$ and $HG_{01}$ modes for each polarization can be reduced to the following compact form

\begin{equation}
\delta\Phi^{i} = 2\arctan\left(\frac{\dfrac{z_{Rx}^{i}}{f_{x} - l_{1,x}^{i}} - \dfrac{z_{Ry}^{i}}{f_{y} - l_{1,y}^{i}}}{1 + \dfrac{z_{Rx}^{i} z_{Ry}^{i}}{(f_{x} - l_{1,x}^{i})(f_{y} - l_{1,y}^{i})}}\right), 
\label{eq:gouy_one_order}
\end{equation}
as shown in Eq.~\eqref{eq:gouy_simplify}. Notice that the phase difference $\delta\Phi^{i}$ depend exclusively on the resonator geometry, while the cavity linewidth is defined by the reflectivity of the cavity mirrors.

\subsection{Conditions for folded cavity}
As seen in Sec.~\ref{Sec:condition_semi}, a semimonolithic cavity with curvature radius $R = 25$ mm and the arm length $l_1$ = 20 mm for $l_c=5$ mm presents an accentuated astigmatim for the ordinary polarization. We use a folded cavity to compensate this astigmatim for both polarizations. The folded cavity has several degrees of freedom $l_1$, $l_2$, $\theta$, and $R$ so it was necessary to restrict the parameter space to maintain experimental control. The mirror curvature radius $R = 25$ mm and the arm length $l_1$ = 20 mm were kept fixed, with only the degree of freedom $\theta$ varying. Additionally, the condition $l_{2}$ $\approx$ 0.9 $l_{1}$ was imposed to preserve the mode-matching conditions between the chosen astigmatic linear cavity and the folded cavity, avoiding the need to modify the input optics.

\begin{figure}[h]
\centering
\includegraphics[width=0.49\textwidth]{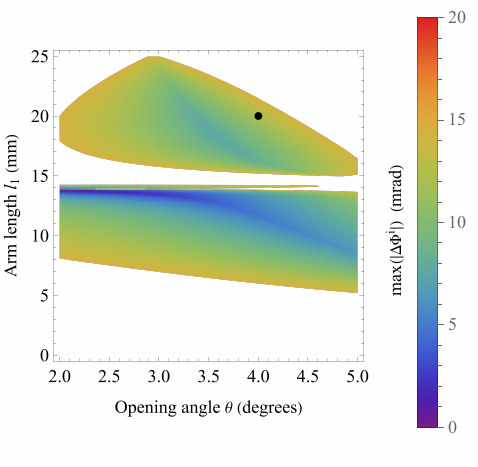}
\caption{(Color online) Parameter space $\left(l_{1} \times \theta \right)$ for the folded cavity
with $l_2 = 0.9\,l_1$ and $R = 25\,\mathrm{mm}$. The color scale indicates the larger
of the ordinary and extraordinary phase separations, in mrad, i.e.,
$\max \left(\left|\delta\Phi^{i}\right|\right)$, with $i = o,e$. The white region corresponds
to pairs $\left(l_{1},\,\theta\right)$ for which at least one of the separations
exceeds 20 mrad, the linewidth of the resonator. Within the colored region,
both polarizations simultaneously satisfy $\left|\delta\Phi^{i}\right| \leq 20$ mrad.
The black dot marks the configuration used in the experimental setup.}
\label{fig:espaco_parametro}
\end{figure}

Fig.~\ref{fig:espaco_parametro} illustrates the parameter space $\left(l_{1} \times \theta\right)$, where the colored region delimits the pairs $\left(l_{1},\,\theta\right)$ for which $\delta \Phi ^{i}$ simultaneously satisfy the compensation condition of Eq.~\eqref{eq:condition_resonance}. Due to crystal birefringence, the ordinary and extraordinary polarizations experience distinct effective optical path lengths, yielding different round-trip transverse phase shifts ($\delta \Phi^{o}$ and $\delta \Phi^{e}$). For simultaneous resonance, both phase shifts must remain below the 20~mrad cavity linewidth, represented by the parameter $\max(\vert{}\delta\Phi^{i}\vert{})$. In the color map, violet-to-blue regions denote configurations well below this threshold, orange-to-red areas approach the limit, and white regions indicate that at least one polarization exceeds it. The marked point ($l_1 = 20$~mm, $\theta = 4^\circ$) corresponds to the configuration used in our experimental setup, for which $\delta\Phi^{o} = 8.30$~mrad and $\delta\Phi^{e} = 17.40$~mrad. Although smaller phase shifts are theoretically attainable (blue bands), this operating point was selected to accommodate physical experimental constraints.

\section{EXPERIMENTAL RESULTS}\label{Sec:exp_res}
In this section, we present the experimental results obtained from three cavity configurations: the semimonolithic cavity with $R$ = 25 mm, the folded cavity with $R$ = 25 mm, and the semimonolithic cavity with $R$ = 2000 mm. First, we aim to demonstrate that a beam carrying OAM exhibits separation between the first-order modes due to the crystal birefringence in a high-finesse cavity. The second result supports the theoretical proposal that a folded cavity is capable of minimizing astigmatism while preserving the angular momentum in the ordinary and extraordinary polarizations of the crystal. The third result obtained with the semimonolithic cavity of $R$ = 2000 mm, demonstrates that, depending on the radius of curvature of the resonator, the OAM is preserved in the polarizations of the birrefringent medium. 

\begin{figure}[h]
\centering
\includegraphics[width=0.495\textwidth]{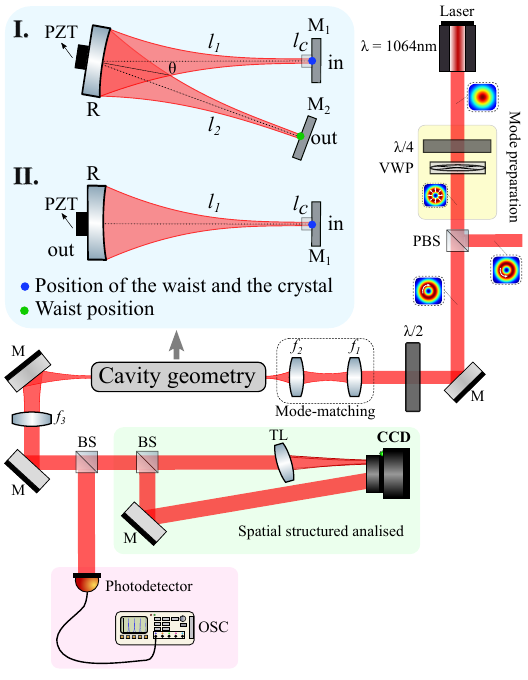} \caption{(Color online) Experimental setup for cavity-enhanced experiments with a birrefringent medium. A CW laser at 1064 nm is used as the light source. The mode preparation stage consists of a quarter-wave plate (QWP) and vortex wave plate (VWP), followed by a polarizing beam splitter (PBS) to project the LG mode. Flat mirrors (M) are used for beam steering. A half-wave plate (HWP) placed before the mode-matching system simulates the converted signal and idler modes. The two-lens system ($f_1$ and $f_2$) focuses the input beam to the requiredwaist at mirror $M_{1}$ for both cavity geometries: \textbf{I.} folded cavity and \textbf{II.} semimonolithic cavity. The geometry of each cavity is characterized by the flat mirrors $M_{1}$ and $M_{2}$, the arm lengths $l_{1}$ and $l_{2}$, and the crystal length $l_{c}$. A lens $f_{3}$ collimates the output beam from the cavity. The detection system consists of: spectral detection using a photodetector and an oscilloscope (OSC); and spatial mode tomography with two optical paths one through a tilted lens (TL) and one direct path enabling simultaneous detection of the OAM mode and topological charge.}\label{fig:setup}
\end{figure}

The experimental setup is depicted in Fig.~\ref{fig:setup}. The light source is a continuous-wave (CW) Nd:YAG laser (Innolight, model \textit{Diabolo} 1000 NE) operating at 1064 nm with an output power of 250 mW. Prior to mode preparation, the beam passes through a Faraday isolator, which prevents optical feedback into the laser cavity, followed by a spatial filtering system based on a pinhole, which ensures the spatial quality of the input beam into the cavity (not shown in Fig.~\ref{fig:setup}).

In all experiments, as shown in Fig.~\ref{fig:setup}, a first-order Laguerre-Gauss (LG) mode is prepared and injected into the cavity.
The nonlinear medium used in all cavity configurations is a Type II KTP crystal, with length $\textit{l}_c$ = 5 mm, antireflection (AR) coatings on both faces for 532 nm and 1064 nm, refractive indices $n_x$ = 1.7404, $n_y$ = 1.7479, and $n_z$ = 1.8296 at 1064 nm, and an extraordinary refractive index of $n_e$ = 1.7467 at 1064 nm. The crystal was placed on the inner face of the input mirror $M_1$ in all cavity configurations.
The cavity geometries are highlighted in Fig.~\ref{fig:setup}: \textbf{I.} the folded cavity and \textbf{II.}, the semimonolithic cavity, both to piezoeletric transducer (PZT) for scanning. The semimonolithic cavity includes an output port responsible for the scanning signal. Depending on the chosen cavity geometry, a lens $f_{3}$ collimates the output beam onto a 50/50 beam spitter (BS), where part of the signal is detected by a photodetector Thorlabs PDA100A2 and analyzed with a Tektronix DPO2014B oscilloscope (OSC). The remaining portion of the signal is directed into a spatial mode tomography system, where a second BS splits the beam: one part passes through a tilted lens (TL) positioned at a distance equal to its focal length from a CCD camera, enabling visualization of the mode conversion; the other part propagates without any optical conversion. This detection scheme is used to perform tomography of structured modes \cite{ PhysRevA.103.063704,48bj-bm8b}.

The transmission signals were acquired using a photodetector and displayed on a oscilloscope. Given the high reflectivity of the cavity mirrors, the resonance peaks of interest were well-approximated and fitted with Lorentzian functions to extract the relevant parameters, namely: linewidth, finesse, and phase separation between the modes $\delta \Phi^{i}_{j}$, for both the ordinary and extraordinary polarizations.

\subsection{Semimonolithic Cavity R = 25 mm}\label{sub:semi_25}

The semimonolithic cavity was assembled using a high-reflectivity (HR) curved mirror with radius of curvature $R$ = 25 mm, manufactured by Laseroptik, with reflectivity $\mathcal{R}$ $>$ 99.7\% at 1064 nm. The flat mirror $M_1$ was configured for HR 1064 nm at normal incidence (Type II), with $\mathcal{R}$ $>$ 99.7\%. The mirror separation was set to 20 mm, and the mode-matching waist at the crystal position was calculated to be $w_0$ = 47 $\mu$m. The cavity was operated in scanning mode by applying a ramp signal to a piezoeletric (PZT) coupled to the curved mirror, allowing the longitudinal modes to be mapped as a function of cavity length variation.

\begin{figure}[h]
\centering
\includegraphics[width=0.495\textwidth]{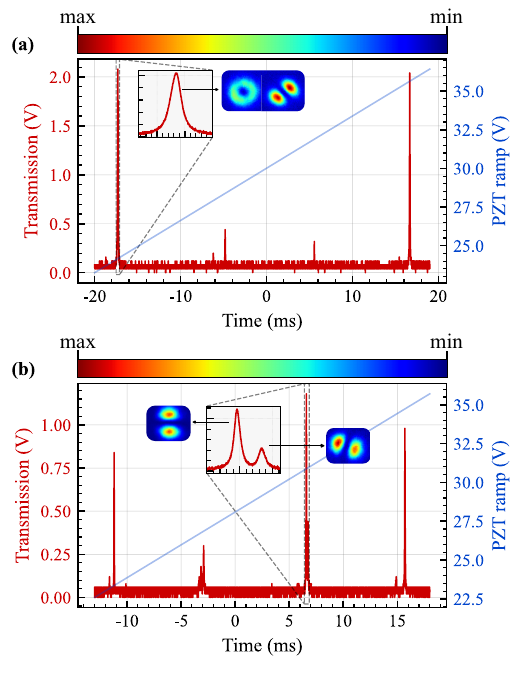} \caption{\label{fig:semi_25mm} (Color online) Transmission spectrum of the semimonolithic cavity. The curves show the resonance signal (red) and the ramp signal applied to the PZT (blue). \textbf{(a)} extraordinary polarization (horizontal): a single resonance peak is observed for the $LG_{0}^{\pm1}$ mode, confirming that the phase separation between the $HG_{10}$ and $HG_{01}$ modes is smaller than the cavity linewidth, on the side is the spatial profile of the $LG_{0}^{\pm1}$ mode with its topological charge signature and the first-order diagonal HG mode resulting from the tilted lens transformation. \textbf{(b)} ordinary polarization (vertical): the resonance peak splits into two, evidencing the degeneracy breaking between the $HG_{10}$ and $HG_{01}$ components of the pump $LG_{0}^{\pm1}$ mode; next to each peak an inset shows the spatial profile of the $HG_{10}$ and $HG_{01}$ components.}
\end{figure}

As shown in Fig.~\ref{fig:semi_25mm}, no splitting is observed in the transmission peaks for the extraordinary polarization (horizontal) \textbf{(a)}. In contrast, a longitudinal splitting is observed for the ordinary polarization (vertical) \textbf{(b)}, evidencing the separation between the first-order LG mode components induced by the birefringence of the KTP Type II crystal. These results confirm that the crystal birefringence acts selectively on the cavity mode structure depending on the polarization state.

The experimental data for this cavity geometry are presented in Table ~\ref{tab:semi_25mm}.

\begin{table}[h]
\caption{\label{tab:semi_25mm}%
 Resonance parameters of the semimonolithic cavity R = 25 mm for ordinary and extraordinary polarizations. }
\begin{ruledtabular}
\begin{tabular}{cccc}
\multicolumn{1}{c}\textrm{Polarization}&\multicolumn{1}{c}{$\delta \phi$}(mrad)&\multicolumn{1}{c}{$\delta \Phi^{teo}$}(mrad)&
\multicolumn{1}{c}{\textrm{Finesse}}\\
\hline
\mbox{extraordinary}&6.80&1.84&461.30\\
\mbox{ordinary}&3.34&23.52&939.01
\end{tabular}
\end{ruledtabular}
\end{table}

The theoretical values of the phase separation between the $HG_{10}$ and $HG_{01}$ modes for this cavity geometry are 1.84 mrad for the extraordinary polarization and 23.52 mrad for the ordinary polarization. For the extraordinary polarization, the linewidth was experimentally measured from the $LG_{0}^{\pm1}$ mode, as it was the lowest-order mode obtained in the cavity scan, yielding 6.8 mrad. This value exceeds the theoretical mode separation, indicating that the phase separation is smaller than the linewidth for this polarization, which manifests as a single resonance peak in the transmission spectrum, confirming the preservation of OAM. In contrast, for the ordinary polarization, the linewidth was measured from the fundamental $HG_{00}$ mode, also obtained from the cavity scan, yielding 3.34 mrad. The experimental separation between the $HG_{10}$ and $HG_{01}$ modes was 17.34 mrad significantly larger than the linewidth evidencing two distinct resonance peaks in the transmission spectrum. This discrepancy prevents OAM conservation in this polarization, as shown in Fig.~\ref{fig:semi_25mm} \textbf{(b)}. 

\subsection{Folded Cavity R=25mm}\label{sub:dobrada_25}

The result of Sec.~\ref{sub:semi_25} demonstrates that, for a high-finesse semimonolithic cavity with $R$ = 25 mm, OAM conservation is inhibited in the ordinary polarization due to the mode splitting induced by the crystal birefringence. To overcome this limitation, we investigate a folded cavity geometry designed to compensate for the astigmatism introduced by the crystal, while simultaneously preserving OAM between the pump beam and the converted modes signal and idler.

The folded cavity consists of three mirrors, as shown in Fig.~\ref{fig:folded_cavity}: two flat mirrors $M_1$ and $M_2$ and one curved mirror ($R$), with radius of curvature $R$ = 25 mm. All mirrors were manufactured by Laseroptik with HR 1064 nm configuration at normal incidence, with reflectivity $\mathcal{R}$ $ > $ 99.7\% at 1064 nm. In order to preserve the mode-matching conditions established in the previous configuration, the input waist was maintained at $w_{0}$ = 47 $\mu$m. The cavity geometry was optimized to compensate for the astigmatism induced by the crystal, yielding the parameters: $l_{1}$ = 20 mm, $l_{2}$ = 17.8 mm. The curved mirror occupies the vertex of the V-shaped geometry, forming an angle $\theta = 3.78^\circ$ between its optical axis and each cavity arm, such that the total opening angle of the cavity is 2$\theta$.

\begin{figure}[h]
\centering
\includegraphics[width=0.495\textwidth]{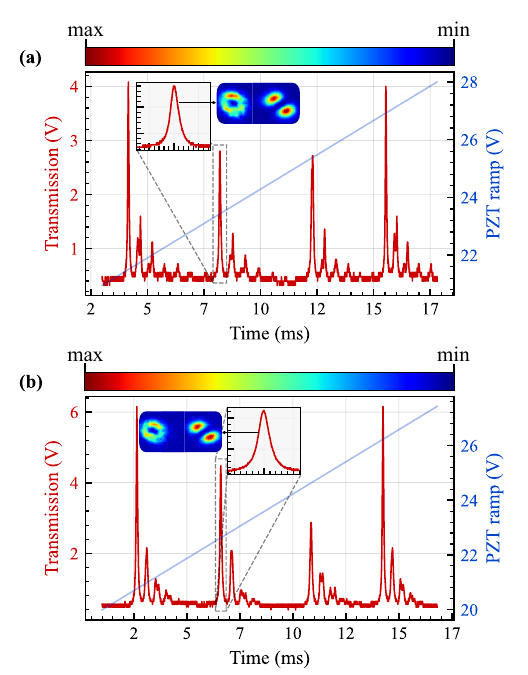} \caption{\label{fig:dobrada_25mm} 
 (Color online) Transmission spectrum of the folded cavity with R = 25 mm, $l_1$ = 20 mm, $l_2$ = 17.8 mm, and $\theta = 3.78$º. The curves show the resonance signal (red) and the modulation applied to the PZT (blue). \textbf{(a)} Extraodinary polarization (horizontal): the inset highlights a single resonance peak for the $LG_{0}^{\pm 1}$ mode, confirming that the phase separation between the $HG_{10}$ and $HG_{01}$ modes (17.4 mrad) is smaller than the cavity linewidth (17.69). The highest amplitude peak corresponds to the fundamental $HG_{00}$ mode, which results from residual coupling due to imperfect mode matching and serves as the linewidth reference. \textbf{(b)} Ordinary polarization (vertical): the inset likewise reveals a single resonance peak, with a theoretical phase separation of 8.34 mrad smaller than the linewidth of 19.77 mrad. The spatial profiles of the $LG_{0}^{\pm1}$ mode and its topological charge signature (first-order diagonal HG mode) are shown alongside each spectrum.}.
\end{figure}

As illustrated in Fig.~\ref{fig:dobrada_25mm}, for the extraordinary polarization \textbf{(a)} a single resonance peak is observed for the first-order mode, and the same behavior is reproduced for the ordinary polarization \textbf{(b)}, indicating that the phase separation between the $HG_{10}$ and $HG_{01}$ modes remained below the cavity linewidth in both polarizations. These results demonstrate that the folded geometry effectively compensated for the astigmatism induced by the crystal, enabling the preservation of OAM  simultaneously in both the ordinary and extraordinary polarizations. 

The corresponding experimental data are presented in Table.~\ref{tab:dobrada_25mm}

\begin{table}[h]
\caption{\label{tab:dobrada_25mm}%
 Resonance parameters of the folded cavity R = 25 mm for ordinary and extraordinary polarizations. }
\begin{ruledtabular}
\begin{tabular}{cccc}
\multicolumn{1}{c}\textrm{Polarization}&\multicolumn{1}{c}{$\delta \phi$}(mrad)&\multicolumn{1}{c}{$\delta \Phi^{teo}$}(mrad)&
\multicolumn{1}{c}{\textrm{Finesse}}\\
\hline
\mbox{extraordinary}&17.69&17.40&177.61\\
\mbox{ordinary}&19.77&8.30&158.89
\end{tabular}
\end{ruledtabular}
\end{table}
The experimental linewidths were measured from the fundamental Gaussian mode $HG_{00}$, yelding 17.69 mrad for the extraordinary polarization and 19.77 mrad for the ordinary polarization. In both cases, the theoretical mode separation is smaller than the experimentally measured linewidth, confirming that the $HG_{10}$ and $HG_{01}$ modes are transmitted within a single resonance peak. It is worth noting that, for the extraordinary polarization, the theoretical separation of 17.4 mrad approaches the limit set by experimental linewidth of 17,69 mrad, which reflects the experimental challenges inherent to this geometry, such as mirror alignment, absorption, and oblique incidence on the curved mirror. As can be seen in Fig.~\ref{fig:dobrada_25mm} \textbf{(a)} and \textbf{(b)}, a single resonance peak is observed for both the extraordinary and ordinary polarizations, demonstrating that the folded cavity geometry enables the simultaneous preservation of OAM in both polarizations.

\subsection{Semimonolithic Cavity $R$ = 2000 mm}\label{sub:semi_2m}

\begin{figure}[h]
\centering
\includegraphics[width=0.495\textwidth]{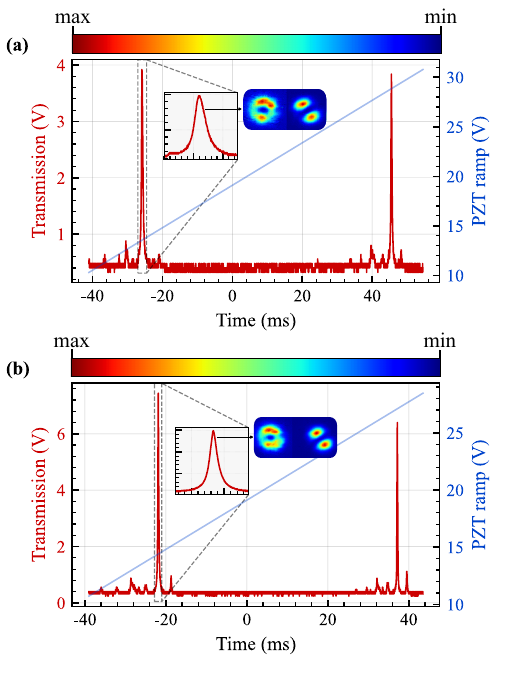} \caption{(Color online) Transmission spectrum of the semimonolithic cavity with $R$ = 2000 mm and $l_{1}$ = 46 mm. The red curve shows the signal recorded by the photodetector and the blue curve is the PZT modulation signal. (\textbf{a}) Extraordinary polarization (horizontal): the inset shows a single resonance peak for the $LG_{0}^{\pm1}$ mode, confirming that the phase separation remains below the cavity linewidth theoretical value $\delta\Phi^{e}$ = 0.07 mrad, experimental linewidth $\delta\phi^{e}$ = 21.00 mrad, on the side is the spatial profile of the $LG_{0}^{\pm1}$ mode and its topological charge signature, a first-order diagonal HG mode resulting from the tilted lens transformation. (\textbf{b}) Ordinary polarization (vertical): the inset likewise shows a single resonance peak for the $LG_{0}^{\pm 1}$ mode theoretical value $\delta\Phi^{o}$ = 0.90 mrad, experimental linewidth $\delta\phi^{o}$ = 12.07 mrad, on the side is the spatial profile of the $LG_{0}^{\pm1}$ mode and the first-order diagonal HG mode obtained from the tilted lens.  Note: Each spatial profile was independently normalized to its own maximum intensity for visualization purposes.}\label{fig:semi_2m}
\end{figure}

The semimonolithic cavity with $R$ = 2000 mm represents a particular case of resonator geometry that minimizes astigmatic effects on the transverse modes. In this configuration, mode selection is no longer governed by the crystal birefringence, but rather by the geometric effect of the mirror radius of curvature which, being sufficiently large, results in a very small Gouy phase accumulated per round trip, such that the separation between first order modes falls well below the cavity linewidth.This allows for sustained oscillation of OAM modes inside the cavity with both ordinary and extraordinary polarizations, as shown in Fig. \ref{fig:semi_2m}. The distance between the curved mirror and the flat mirror $M_1$ is $l_{1}$ = 46 mm, with a crystal length of $l_{\text{c}}$ = 5 mm a mode-matching waist of $w_{0}$ = 315 $\mu$m. The curved mirror has reflectivity $\mathcal{R} > 99.7\%$ at 1064 nm, while the flat mirror $M_1$ was configured for HR 1064 nm, with $\mathcal{R} > 99.7\%$ at 1064 nm. The Type II KTP crystal, which has white AR coatings on both faces, was positioned in the vicinity of the flat mirror, analogously to the previous cavity configurations.

\begin{table}[h]
\caption{\label{tab:semi_2000mm}%
 Resonance parameters of the semimonolithic cavity R = 2000 mm for ordinary and extraordinary polarizations. }
\begin{ruledtabular}
\begin{tabular}{cccc}
\multicolumn{1}{c}\textrm{Polarization}&\multicolumn{1}{c}{$\delta \phi$}(mrad)&\multicolumn{1}{c}{$\delta \Phi^{teo}$} (mrad)&
\multicolumn{1}{c}{\textrm{Finesse}}\\
\hline
\mbox{extraordinary}&21.00&0.07&149.70\\
\mbox{ordinary}&12.07&0.90&260.30
\end{tabular}
\end{ruledtabular}
\end{table}
The experimental data compiled in Table.~\ref{tab:semi_2000mm} show that the measured phase linewidths of $\delta\phi^{o}$ = 12.07 mrad and $\delta\phi^{e}$ = 21 mrad are significantly greater than the theoretical phase separation of $\delta\Phi^{o}$ = 0.90 mrad and $\delta\Phi^{e}$ = 0.07 mrad, confirming that both first-order mode components are transmitted within a single resonance peak for each polarization. This result is consistent with the theoretical prediction of Sec.~\ref{Sec:condition_semi}: as the resonator is designed with a sufficiently large radius of curvature, the Gouy phase separation decreases continuously, falling well below the cavity linewidth and enabling the simultaneous preservation of OAM between the pump beam and the converted signal and idler modes without the need to fold the cavity.

\section{CONCLUSION}\label{Sec:conclusion}

We have experimentally demonstrated that the astigmatism induced by a birefringent crystal in resonators injected with first-order OAM modes can be effectively compensated using a folded cavity configuration. The folded cavity constitutes a viable alternative when modifying the radius of curvature of the resonator—in this case, a semimonolithic cavity—is not an option: by adding a third mirror and folding the cavity, the mode-matching parameters are preserved, while the condition that the phase separation between the modes remains below the cavity linewidth is simultaneously imposed. Under this condition, the resonator preserves simultaneous resonance for both polarizations of the Type-II KTP crystal without requiring additional intracavity optics.

From an analytical perspective, this phase separation requirement can also be analyzed by directly modifying the mirror's radius of curvature $R$. For sufficiently large values of $R$, the astigmatism arising from the crystal birefringence becomes negligible, naturally keeping the mode phase separation within the cavity linewidth. However, when such geometrical constraints cannot be implemented in a monolithic or semimonolithic setup, the folded cavity provides the necessary control parameter to compensate for the astigmatism. It should be emphasized that an increase in the radius of curvature $R$ leads to an increase in the Rayleigh length of the resonant beam. Consequently, the Boyd condition for maximum nonlinear efficiency—where the Rayleigh length is on the order of the crystal length \cite{boyd_parametric_1968}—is no longer satisfied. This departure reduces the efficiency of the spontaneous process, although it brings the system into the typically employed thin-crystal regime. Notably, for a stimulated nonlinear process, this drop in efficiency can be experimentally tolerated.

Furthermore, the theoretical and experimental analysis presented here focused on a Type-II nonlinear process, in which the converted beams have perpendicular polarizations. However, in Type-0 or Type-I processes (where the polarizations are parallel), the astigmatic correction condition in Eq. \eqref{eq:condition_resonance} is less restrictive, as it needs to be satisfied only for the desired polarization. Under these conditions, it is possible to choose an opening angle $\theta$ for which the astigmatism is perfectly compensated, making the astigmatic cavity an ideal solution.

\section*{ACKNOWLEDGMENTS}
This work was funded by the Brazilian research funding agency, Coordenação de Aperfeiçoamento de Pessoal de Nível Superior (CAPES - Finance Code 001), Conselho Nacional de Desenvolvimento Científico e Tecnológico (CNPq),  Fundação Carlos Chagas Filho de Amparo à
Pesquisa do Estado do Rio de Janeiro (FAPERJ), and Fundação de Amparo
à Pesquisa do Estado de São Paulo (FAPESP, grant 2021/06823-5).

\appendix

\section{Accumulated Gouy phase in the folded cavity}\label{Sec:app_A}

We derive the accumulated Gouy phase in the folded cavity by first analyzing the phase shift along a given transverse direction, neglecting birefringence-induced astigmatism (Sec.~\ref{APsub:1}). Next, we discuss how the effective cavity length is modified by the presence of the birefringent crystal (Sec.~\ref{APsub:2}). Finally, we combine these results to obtain the total Gouy phase for first-order modes (Sec.~\ref{APsub:3}).

\subsection{The Gouy phase in terms of the focal lengths of the resonator}\label{APsub:1}

\begin{figure}[htbp]
  \centering
  \resizebox{8.87cm}{!}{
    \begin{tikzpicture}[x=0.75pt,y=0.75pt,yscale=-1,xscale=1]
    \tikzset{every picture/.style={line width=0.75pt}}
    
    %Straight Lines
    \draw [line width=1.5] (328.47,71.67) -- (328.19,115) -- 
        (327.55,212.83) -- (327.53,216);
    \draw [shift={(327.5,220)}, rotate=270.38] 
        [fill={rgb,255:red,0;green,0;blue,0}][line width=0.08] 
        [draw opacity=0] (13.4,-6.43) -- (0,0) -- (13.4,6.44) -- (8.9,0) -- cycle;
    \draw [shift={(328.5,67.67)}, rotate=90.38] 
        [fill={rgb,255:red,0;green,0;blue,0}][line width=0.08] 
        [draw opacity=0] (13.4,-6.43) -- (0,0) -- (13.4,6.44) -- (8.9,0) -- cycle;
    
    %Bezier Curves
    \draw [color={rgb,255:red,126;green,211;blue,33},draw opacity=1][line width=1.5]
        (470,154.51) .. controls (420.96,154.18) and (355.11,167.25) .. (327.5,220);
    
    %Dashed Line
    \draw [dash pattern={on 0.84pt off 2.51pt}] (190,144.5) -- (470,144.5);
    
    %Curve Lines
    \draw [color={rgb,255:red,126;green,211;blue,33},draw opacity=1][line width=1.5]
        (190,134.5) .. controls (221.41,134.72) and (259.91,129.01) .. 
        (290.19,109.68) .. controls (305.87,99.68) and (319.35,86.03) .. (328.5,67.67);
    
    \draw [color={rgb,255:red,126;green,211;blue,33},draw opacity=1][line width=1.5]
        (190,154.51) .. controls (237.32,154.18) and (300.86,167.25) .. (327.5,220);
    
    \draw [color={rgb,255:red,126;green,211;blue,33},draw opacity=1][line width=1.5]
        (470,135.67) .. controls (421.3,136) and (355.91,122.67) .. (328.5,68.84);
    
    %Straight Lines
    \draw (470,135.67) -- (470,154.51);
    \draw (190,134.5) -- (190,154.51);
    
    %Text Nodes
    \draw (400,85) node [anchor=north west][inner sep=0.75pt] {$z'_{R}$};
    \draw (245,90) node [anchor=north west][inner sep=0.75pt] {$z_{R}$};
    \draw (322,48) node [anchor=north west][inner sep=0.75pt] {$f$};
    \draw (369,144.67) node [anchor=north west][inner sep=0.75pt] {$l_{2}$};
    \draw (269,144.67) node [anchor=north west][inner sep=0.75pt] {$l_{1}$};
    
    \end{tikzpicture}
  }
  \caption{Rayleigh length mapping by the curved mirror in the folded cavity.}
  \label{fig:diagrama_zr}
\end{figure}
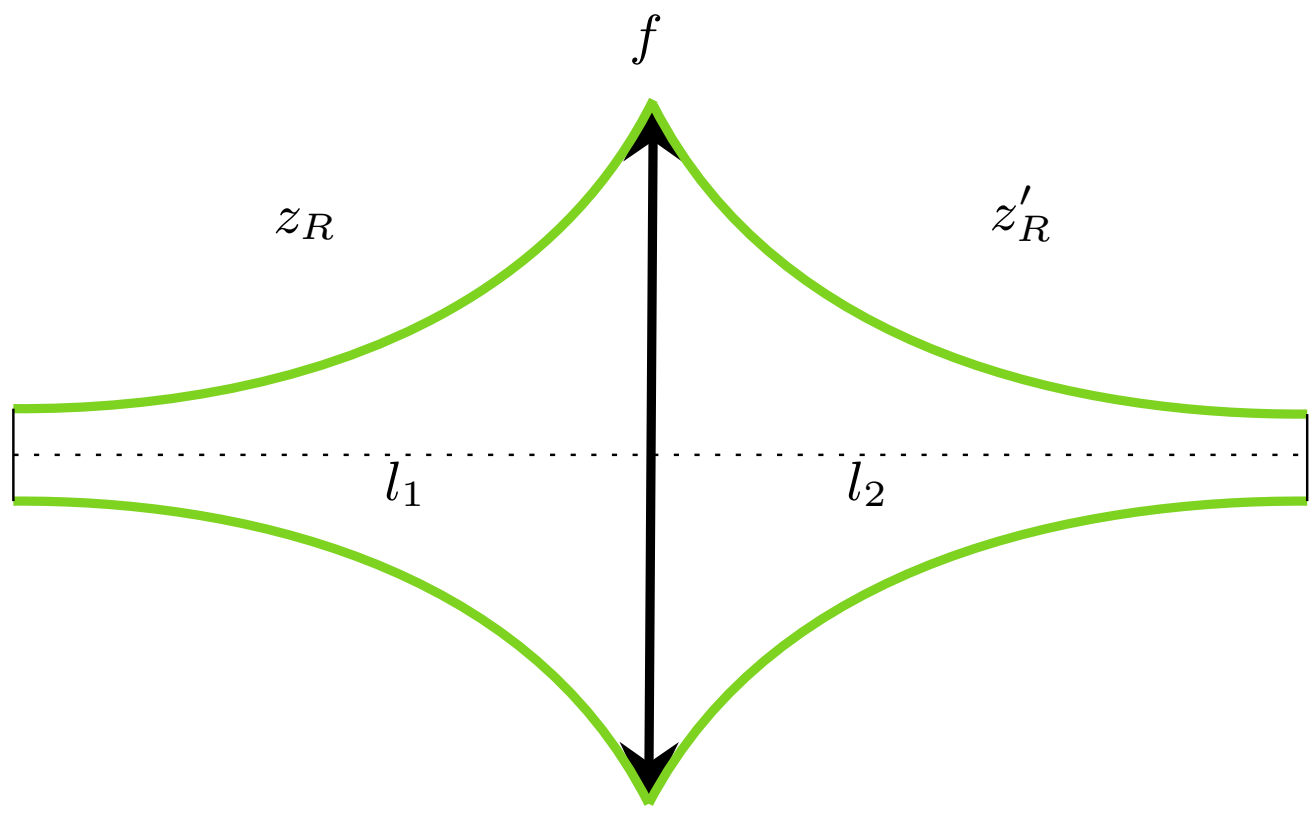

In this appendix, we present a simplified derivation of the Gouy phase accumulated per round trip in the folded cavity. Since the folded cavity has two foci, both located at the surfaces of the flat mirrors $M_1$ and $M_2$, the problem reduces to analyzing the beam propagation along each cavity arm independently. The curved mirror is equivalently represented by a lens $f$, which transforms the beam Rayleigh length from $z_R$ to $z'_R$ and shifts its focal position from $l_1$ to $l_2$ relative to the lens. This configuration is illustrated in Fig.~\ref{fig:diagrama_zr}.

The geometric constraint between the two cavity arms is that the cavity lenght $L = (l_{1} + l_{2})$. For the fundamental Gaussian mode the Gouy phase accumulated during a round-trip in the optical cavity is the sum of the contributions from each arm:

\begin{equation}
    \Phi = 2\arctan\left(\frac{l_{1}}{z_R}\right) + 2\arctan\left(\frac{l_{2}}{z'_R}\right),
\end{equation}

where $z_R$ and $z'_R$ are the Rayleigh lengths associated with each cavity arm, which are in general distinct since $l_{1}$ $\ne$ $l_{2}$. Applying the trigonometric identity:

\begin{equation}
    \arctan\left(x\right) + \arctan\left(y\right) = \arctan\left(\frac{x+y}{1-xy}\right),
\end{equation}

with $x = l_{1}/z_R$ and $y = l_{2}/z'_R$, the accumulated Gouy phase can be written in compact form as:

\begin{equation}
    \Phi = 2\arctan\left(\frac{z'_R \,l_{1} + z_R \,l_{2}}{z'_R \,z_R - l_{1}\,l_{2}}\right).\label{eq:A3}
\end{equation}

From Fig.~\ref{fig:diagrama_zr}, we can identify the output Rayleigh length $z_{R}'$ and the distance $l_{2}$ as:

\begin{equation}\label{eq:A6}
    z'_R = \frac{f^{2}z_{R}}{\left(l_{1} - f\right)^2 +z^2_{R}},
\end{equation}

\begin{equation}\label{eq:A7}
    l_{2} = f + \frac{\left(l_{1} - f\right)f^{2}}{\left(l_{1} - f\right)^2 + z_{R}^2}\,.
\end{equation}

Substituting Eqs.\eqref{eq:A6} and \eqref{eq:A7} into Eq.\eqref{eq:A3}, one obtains the following result

\begin{equation}\label{eq:A10}
    \Phi = 2\arctan\left(\frac{z_{R}}{f - l_1}\right).
\end{equation}

This remarkably simple result shows that the accumulated Gouy phase depends only on the Rayleigh length $z_{R}$, the equivalent focal length $f$ of the curved mirror, and the arm length $l_{1}$, being explicitly independent of $l_{2}$. However, as shown below in Sec.~\ref{APsub:2}, the propagation of a paraxial beam in a birefringent medium introduces a polarization-dependent astigmatism, making the effective cavity length dependent on the polarization of the propagating field.

\subsection{Astigmatic propagation of a paraxial beam in a birrefringent medium}\label{APsub:2}

Reference \cite{martinelli_orbital_2004} presents a theoretical formulation to describe the paraxial propagation of optical beams in an anisotropic medium, extending the modeling of uniaxial crystals to the case of biaxial crystals—specifically KTP (which exhibits quasi-uniaxial behavior, with $n_x \sim n_y \neq n_z$) placed inside an optical cavity.
As it propagates through the biaxial crystal (KTP), the beam becomes astigmatic. The treatment is carried out in rotated coordinates $\left(x', y', z\right)$, with $x'$  along the propagation direction, forming an angle $\phi$ with the crystallographic axis $x'$; $\phi$ is the crystal's cut angle, done during manufacturing \cite{martinelli_orbital_2004}. In this geometry, the extraordinary wave (polarized in the $xy$ plane) has a refractive index that depends on the propagation direction, given by the index ellipsoid projected onto the $xy$ plane \cite{martinelli_orbital_2004, yariv1989quantum}:
\begin{equation}
    \frac{1}{n^2} = \frac{\cos^2\phi}{n_y^2} + \frac{\sin^2\phi}{n_x^2}.\label{eq:app_1}
\end{equation}
For convenience, let's define $n \equiv n(\phi)$. Using the approximation that KTP is quasiuniaxial ($n_{x} \approx n_{y} \neq n_{z}$), one can obtain a paraxial equation for each polarization. 

For the ordinary wave ($z$-polarization), applying the paraxial approximation $E_z = u_z(x', y', z) e^{-in_z k_0 x'}$ and separating the transverse variables such that $u_z(x', y', z) = U_z(x^\prime,y^\prime) V_z(x^\prime,z)$, yields a pair of paraxial diffraction equations for each propagation plane:
\begin{align}
    \partial^2_{y'}{U}_{z} = 2in_{z}k_{0}\partial_{x'}{U}_{z}, \qquad \frac{n_{z}^{2}}{n^{2}}\partial^{2}_{z}{V}_{z} = 2in_{z}k_{0}\partial_{x'}{V}_{z}\,.\label{eq:dif_pl1}
\end{align}
%and,

For the extraordinary wave ($xy$-plane polarization), adopting the paraxial approximation $E_y = u_y(x', y', z) e^{-in k_0 x'}$, the propagation equation includes terms originating from the walk-off angle $\phi'$ (the angular separation between the wavevector and the Poynting vector). To absorb this effect, a second coordinate transformation is applied ($y'' = y' - x' \tan\phi'$), corresponding to a transverse shift of the beam. Factorizing the envelope such that $u_y(x^\prime,y^{\prime},z) = U_y(x^\prime,y^{\prime\prime}) V_y(x^\prime,z)$ results in two new paraxial equations for each transverse propagation plane:
\begin{align}
    \xi^2\partial_{y''}^2{U}_y = 2ik_0\frac{n_y^2}{n}\partial_{x'}{U}_y, \qquad \partial_z^2{V}_y = 2ik_0\frac{n_y^2}{n}\partial_{x'}{V}_y\,,\label{eq:dif_pl2}
\end{align}
where $\xi^{2} = \sin^2 \phi + (n^2_{y}/n^2_{x})\cos^2{\phi}$. 
Each of these four equations, Eqs.~\eqref{eq:dif_pl1} and \eqref{eq:dif_pl2}, has the standard paraxial diffraction form, where $\Psi$ generically denotes either amplitude $U$ or $V$, and $u$ is the transverse coordinate of the correponding plane ($y', y'', \text{or}\,z$):

\begin{equation}
    \partial^{2}_{u}\Psi = 2ik_{\text{eff}}\,\partial_{x'} \Psi,\label{eq:paraxial_crys}
\end{equation}

with,

\begin{equation}
    k_{\text{eff}} = n_{\text{eff}}\,k_{0}\label{eq:app_2},
\end{equation}

where $k_0 = \omega/c$ is the vacuum wavenumber at frequency $\omega$.

From Eq.~\eqref{eq:paraxial_crys}, four effective refractive indices arise - one per transverse plane for each polarization ($o, e$) given by:

\begin{align}
    n^{o}_{\text{eff},y} &= n_{z}, \quad 
    n^{o}_{\text{eff,z}} = \frac{n^{2}}{n_{z}}, \label{eq:eff_1}
    \end{align}
    
    \begin{align}
    n^{e}_{\text{eff},y} &= \frac{n^{2}_{y}}{n \, \xi^{2}}, \quad 
    n^{e}_{\text{eff},z} = \frac{n^{2}_{y}}{n}.\label{eq:eff_2}
\end{align}

This set four effective refractive indices, $n_{\text{eff}}$ described in Eqs.~\eqref{eq:eff_1} and \eqref{eq:eff_2}, was used in the numerical simulation to determine the effective optical cavity lengths along each transverse direction. These lengths are required for the correct calculation of the Gouy phase of the cavity eigenmodes.

When a birefringent crystal is placed inside the cavity, the optical path length experienced by the beam is modified according to the refractive index of the medium. Since the phase velocity of light inside a medium of refractive index $n$ differs from that in free space, the geometric length of the cavity must be replaced by an effective optical path length. For a cavity of geometric length $l$ containing a crystal of length $l_c$ and refractive index $n_{\text{eff}}$, the effective optical path length is given by \cite{martinelli_orbital_2004}:

\begin{equation}\label{eq:length_eff}
l^{i}_{j} = l + l_{c}\left(\frac{1}{n_{\text{eff},j}^{i}}-1\right),
\end{equation}

where $i=\{o,e\}$, and $j$ represents the transverse directions of the crystal input plane. For an input crystal face in the $yz$-plane then $j=\{y,z\}$ and $n_{\text{eff},j}^{i}$ is described by Eqs.~\eqref{eq:eff_1} and \eqref{eq:eff_2}, which represent the effective refractive index of the medium, and generally depends on the polarization of the beam and the orientation of the crystal axes, namely: ordinary polarization (o) and extraordinary polarization (e).
\subsection{Compact form of the Gouy phase for first-order modes}\label{APsub:3}
%Given that the Gouy phase is expressed as:
%
Considering an astigmatic Hermite-Gauss beam propagating in the $z$-direction along the crystal with its input face coinciding with the $xy$-plane the Gouy phase accumulated during a round-trip in the optical cavity, for a given polarization $i = \{o, e\}$, is expressed by Eq.~\eqref{eq:A10}, with each quantity evaluated in its respective transverse direction:

\begin{equation}
\begin{split}
\Phi^{i}_{m n} &= 2\left(m + \frac{1}{2}\right) \arctan \left( \frac{z_{R,x}^{i}}{f_x - l_{1,x}^{i}} \right) \\ &+ 2\left(n + \frac{1}{2}\right) \arctan \left( \frac{z_{R,y}^{i}}{f_y - l_{1,y}^{i}} \right),
\label{eq:C1}
\end{split}
\end{equation}
where the term $l_{1,j}^{i}$ is the effective length described by Eq.~\eqref{eq:length_eff}, adapted here for the transverse directions $j = (x ,y)$ (instead of $y, z$).

The phase separation between the first-order modes is obtained by computing $\Phi^{i}_{1 0}$ and $\Phi^{i}_{0 1}$ separately. 

The difference between the two modes is therefore:
\begin{equation}
\begin{split}
\delta\Phi^{i} &= \Phi^{i}_{10} - \Phi^{i}_{01} = \\[6pt]&2\arctan \left( \frac{z_{R,x}^{i}}{f_x - l_{1,x}^{i}} \right) - 2\arctan \left( \frac{z_{R,y}^{i}}{f_y - l_{1,y}^{i}} \right).
\end{split} 
\end{equation}
Applying the trigonometric identity $\arctan \left(x\right) - \arctan \left(y\right)=\arctan \left(\frac{x - y}{1 + xy}\right)$, the following compact form is obtained:
\begin{equation}
\delta\Phi^{i} = 2\arctan\left(\frac{\dfrac{z_{R,x}^{i}}{f_x - l_{1,x}^{i}} - \dfrac{z_{R,y}^{i}}{f_y - l_{1,y}^{i}}}{1 + \dfrac{z_{R,x}^{i}, z_{R,y}^{i}}{(f_x - l_{1,x}^{i})(f_y - l_{1,y}^{i})}}\right).
\label{eq:gouy_simplify}
\end{equation}

\nocite{*}

\bibliography{ref}% Produces the bibliography via BibTeX.

\end{document}